\documentclass[pdflatex]{sn-jnl}% Math and Physical Sciences Numbered Reference Style 
\usepackage{graphicx}%
\usepackage{multirow}%
\usepackage{amsmath,amssymb,amsfonts}%
\usepackage{amsthm}%
\usepackage{mathrsfs}%
\usepackage[title]{appendix}%
\usepackage{xcolor}%
\usepackage{color}%
\usepackage{textcomp}%
\usepackage{manyfoot}%
\usepackage{booktabs}%
\usepackage{algorithm}%
\usepackage{algorithmicx}%
\usepackage{algpseudocode}%
\usepackage{listings}%
\usepackage{soul}
\usepackage{hyperref}
\usepackage{fullpage}
\usepackage{bm,txfonts}
\usepackage{graphicx}
\usepackage{lineno}
\usepackage[super]{natbib} % 直接加载 natbib 并设置 super 样式
\nolinenumbers
\def\citep#1{\cite{#1}}

\begin{document}

%\title[$^{210}$Pb $\beta$-spectrum]{High-precision measurement and modelling of the threshold-free $\beta$ spectrum in $^{210}$Pb decay: a new window into the weak interaction}
%\title[$^{210}$Pb $\beta$-spectrum]{Probing weak interaction with threshold-free precision $^{210}$Pb $\beta$ spectrum}
\title[$^{210}$Pb $\beta$-spectrum]{Precision measurement and modelling of the threshold-free $^{210}$Pb $\beta$ spectrum}

\author[1,2]{\fnm{Shuo} \sur{Zhang}}
\author[3]{\fnm{Hao-Ran} \sur{Liu}}
\author*[4,5]{\fnm{Ke} \sur{Han}}\email{ke.han@sjtu.edu.cn}
\author*[6]{\fnm{Xavier} \sur{Mougeot}}\email{xavier.mougeot@cea.fr}
\author*[7]{\fnm{Paul-Antoine} \sur{Hervieux}}\email{paul-antoine.hervieux@ipcms.unistra.fr}

\author[8]{\fnm{Tao} \sur{Sun}}

\author[8]{\fnm{Wen-Tao} \sur{Wu}}

\author[9]{\fnm{Robin} \sur{Cantor}}

\author[2]{\fnm{Jing-Kai} \sur{Xia}}

\author[2]{\fnm{Zhi} \sur{Liu}}

\author[3]{\fnm{Jun-Cheng} \sur{Liang}}

\author[3]{\fnm{Fu-You} \sur{Fan}}

\author[10]{\fnm{Le} \sur{Zhang}}

\author[11]{\fnm{Ming-Yu} \sur{Ge}}

\author[12]{\fnm{Xiao-Peng} \sur{Zhou}}

\author[7]{\fnm{Adrien} \sur{Andoche}}

\affil[1]{\orgname{{Advanced Energy Science and Technology Guangdong Laboratory}}, \orgaddress{\street{1 Henanan Street}, \city{Huizhou}, \postcode{516000}, \state{Guangdong}, \country{China}}}

\affil[2]{\orgdiv{Center for Transformative Science}, \orgname{ShanghaiTech University}, \orgaddress{\street{393 Middle Huaxia Road}, \city{Shanghai}, \postcode{201210}, \country{China}}}

\affil[3]{\orgdiv{Division of Ionizing Radiation}, \orgname{National Institute of Metrology}, \orgaddress{\street{18, Beisanhuandong Road}, \city{Beijing}, \postcode{100029}, \country{China}}}

\affil[4]{\orgdiv{State Key Laboratory of Dark Matter Physics, Key Laboratory for Particle Astrophysics and Cosmology (MoE), School of Physics and Astronomy}, \orgname{Shanghai Jiao Tong University}, \orgaddress{\street{800 Dongchuan Road}, \city{Shanghai}, \postcode{200240}, \country{China}}}
\affil[5]{\orgdiv{Sichuan Research Institute}, \orgname{Shanghai Jiao Tong University}, \orgaddress{\street{366 Hupan North Road}, \city{Chengdu}, \postcode{610218}, \country{China}}}

\affil[6]{\orgdiv{Universit{\'e} Paris-Saclay, CEA, List, Laboratoire National Henri Becquerel (LNE-LNHB)}, \orgaddress{\city{Palaiseau}, \postcode{F-91191}  \country{France}}}

\affil[7]{\orgdiv{Universit{\'e} de Strasbourg, CNRS, Institut de Physique et Chimie des Mat{\'e}riaux de Strasbourg}, \orgaddress{\street{UMR 7504}, \city{Strasbourg}, \postcode{F-67000}, \country{France}}}

\affil[8]{\orgdiv{Shanghai Institute of Microsystem and Information Technology}, \orgname{Chinese Academy of Sciences}, \orgaddress{\street{865 Changning Road}, \city{Shanghai}, \postcode{200050}, \country{China}}}

\affil[9]{\orgname{STAR Cryoelectronics}, \orgaddress{\street{25-A Bisbee Court}, \city{Santa Fe}, \postcode{87508-1338}, \state{New Mexico}, \country{USA}}}

\affil[10]{\orgdiv{School of Physics and Astronomy}, \orgname{Sun Yat-Sen University}, \orgaddress{\street{2 Daxue Road, Tangjia}, \city{Zhuhai}, \postcode{519082}, \country{China}}}

\affil[11]{\orgdiv{Institute of High Energy Physics}, \orgname{Chinese Academy of Sciences}, \orgaddress{\street{19B Yuquan Road}, \city{Beijing}, \postcode{100049}, \country{China}}}

\affil[12]{\orgdiv{School of Physics}, \orgname{Beihang University}, \orgaddress{\street{37 Xueyuan Road}, \city{Beijing}, \postcode{100191}, \country{China}}}

\abstract{
\textbf{Beta decay is a fundamental process that governs nuclear stability and serves as a sensitive probe of the weak interaction and possible physics beyond the Standard Model of particle physics.
However, precise measurements of complete $\beta$ decay spectra, particularly at low energies, remain experimentally and theoretically challenging.
Here we report a high-precision, threshold-free measurement of the full $\beta$ decay spectrum of $^{210}$Pb to excited states of $^{210}$Bi, using a transition-edge sensor (TES)-based micro-calorimeter. 
This approach enables the detection of $\beta$ particle energies from 0 keV up to their endpoint by coincidence summing with subsequent de-excitation energy, thereby eliminating reconstruction artifacts near zero energy that have traditionally limited low-energy spectral accuracy.
To our knowledge, this is the first complete, high-precision $\beta$ decay spectrum from 0 keV.
The data resolve theoretical uncertainties associated with the atomic quantum exchange (AQE) effect.
An accompanying {\it ab initio} theoretical framework, incorporating atomic, leptonic, and nuclear components, predicts a statistically significant (7.2 $\sigma$) enhancement in $\beta$ emission probability near zero energy, in agreement with the measurement and in contrast to models that omit AQE corrections.
These results provide a new benchmark for $\beta$ decay theory at low energies, deepen our understanding of the weak interaction, and establish a critical foundation for searches for new physics, including dark matter interactions and precision studies of neutrinos.}
}

%\keywords{$\beta$ Decay, Transition edge sensor, Atomic quantum exchange, dark matter}

\maketitle

%\section{Introduction}\label{sec1}
Beta decay has been a pivotal probe in fundamental physics for more than a century, from the mystery of its continuous emission spectrum in conflict with the energy conservation principle, confirmed thanks to the very first coincidence measurement a century ago in 1925~\cite{Bot25}, to the discovery of parity violation in the transition~\cite{Lee56,Wu57}, of the neutrino~\cite{Cow56} and its helicity~\cite{Gol58}.
Nowadays, $\beta$ decay is still used to test the Standard Model (SM) of particle physics, such as for the unitarity of the Cabibbo-Kobayashi-Maskawa (CKM) matrix, for the value of its up-down quark mixing element $V_{ud}$, or for the influence of the weak-magnetism induced current~\cite{Har20}. 
A high-precision determination of the complete energy spectrum of the emitted electron is still very challenging.
Such spectra are crucial in radionuclide metrology for realizing the unit of activity in the International System (SI), the becquerel (Bq)~\cite{Kos15,Kos18}, in nuclear medicine radiation dosimetry for advancing precision therapies~\cite{2019-Hashempour-JNMB}, and in nuclear reactors for safeguarding operational safety through decay heat management~\cite{Dor22}.

The $\beta$ spectral shape attracts additional attention due to its impact on new physics searches beyond the SM, particularly at very low energies.
In underground experiments for direct detection of dark matter~\cite{LZ:2022lsv,PandaX:2024qfu,XENON:2023cxc, DarkSide-20k:2017zyg, SuperCDMS:2016wui}, $\beta$ particles emitted from $^{3}$H, $^{39}$Ar, $^{85}$Kr, and $^{210, 212,214}$Pb decays are the primary background sources of the electron recoil (ER) spectrum in the region of interest (ROI).
For example, in liquid xenon detectors~\cite{XENON:2023cxc,PandaX:2024qfu, LZ:2022lsv}, the contribution of $^{214}$Pb, the most essential ER background, depends heavily on the theoretical modelling of the $\beta$ spectrum at very low energy, where their high atomic number induces strong atomic effects on the $\beta$ spectrum.
Significant uncertainties in the theoretical calculations of the low-energy region of these spectra limit the sensitivity for the detection of new physics with ER data~\cite{2020-XENON-PRD, PandaX:2024cic}. 
In addition, the low-energy region of the $\beta$ spectrum is directly correlated to the emission probability of the neutrino in the endpoint energy region, which might help explain the reactor antineutrino spectrum anomaly~\cite{Zhang:2023zif}.

\begin{figure*}[tb]
\centering
\includegraphics[width=1.\hsize]{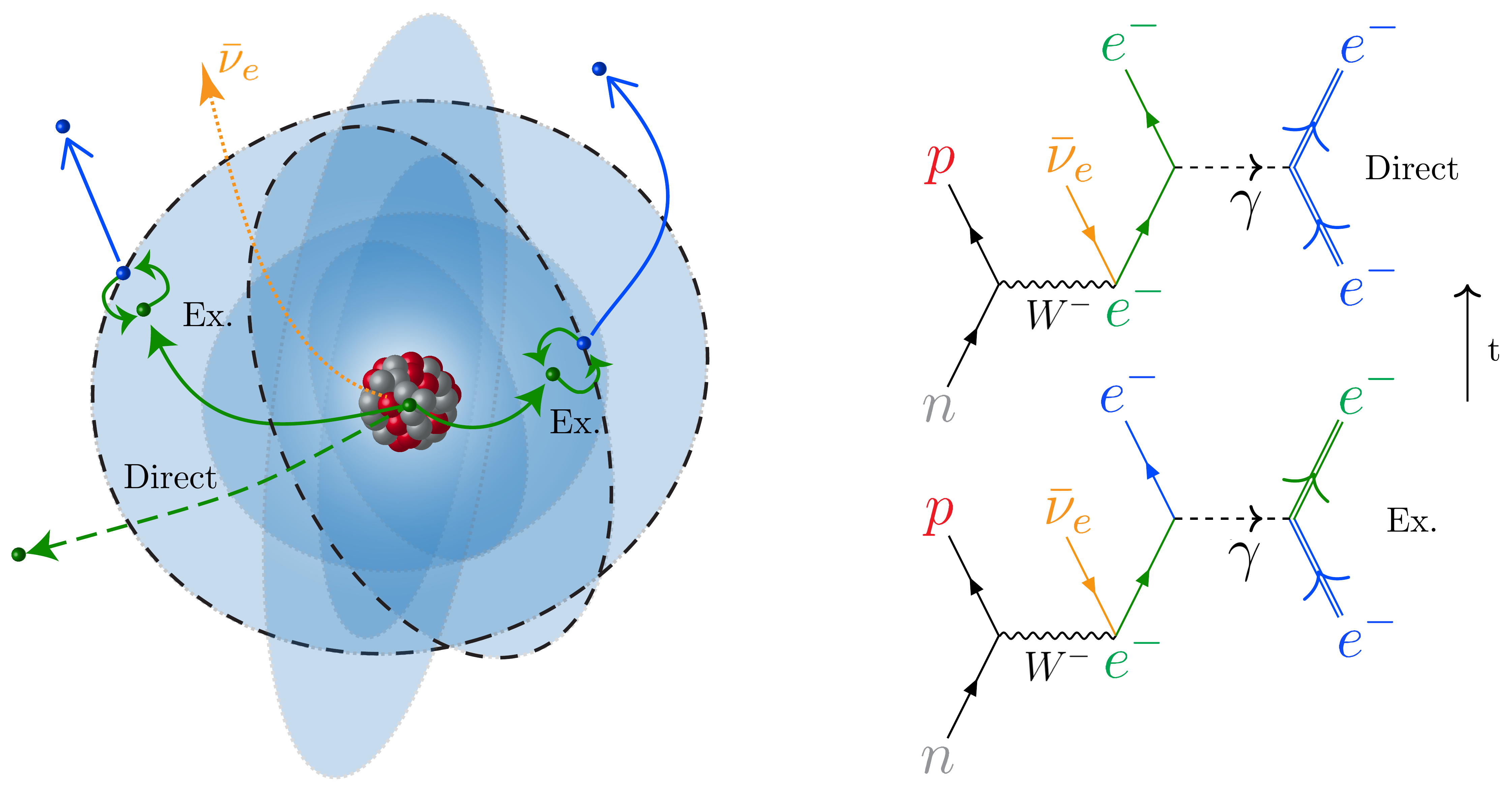}
\caption{Illustration of the atomic quantum exchange (AQE) effect. (Left) In addition to direct decay to the continuum (green dashed line), a $\beta$ particle can be exchanged with an atomic electron (green solid lines) of the parent atom with the same angular momentum. The bound electron is then ejected into the continuum (blue solid lines) and cannot be distinguished from the created $\beta$ particle by any detection system. (Right) Representation of the AQE process in the form of Feynman diagrams. The double lines indicate bound atomic electrons.}
\label{fig:exchange}
\end{figure*}

The decay spectral shape is governed by three fundamental forces -- weak, strong, and electromagnetic interactions -- and depends on the nuclear and lepton wave functions of the parent and daughter atoms. 
While nuclear structure can be of high importance for the whole spectral shape~\cite{Has20}, atomic quantum exchange (AQE) is the most critical effect at very low energy, among many others affecting the $\beta$ emission probability.
This pure quantum effect, depicted in Fig.~\ref{fig:exchange}, refers to the exchange between a $\beta$ particle and an atomic electron: the former is created in an atomic shell, causing the emission of the latter from the same shell of the parent atom~\cite{1963-John-PR}.
AQE, as an additional channel from the usual direct decay to the continuum, enhances the emission probability over the entire spectrum.
The enhancement is especially pronounced at low energies~\cite{1985-Haxton-PRL,1992-Harston-PRA}, where the emission probability can be increased by approximately 20\% below 1 keV. 

Limited measurements are available to test the theoretical predictions of AQE at the lowest energies. 
The reliability of the AQE corrections hinges heavily on the precision of the relativistic wave functions of bound and continuum atomic electrons.
A relatively simple atomic model~\cite{2023-Mougeot-ARI} reproduced $\beta$ spectra of $^{63}$Ni and $^{241}$Pu decays down to the lowest measured energies (0.3 and 0.5 keV, respectively)~\cite{2014-Loidl-63Ni,2010-Loidl-241Pu}. 
However, $^{99}$Tc~\cite{Pau24} and $^{151}$Sm~\cite{2022-Karsten-ARI} measurements revealed that the emission probability exceeds AQE prediction below 5~keV. 
A recent study~\cite{2023-Nitescu-PRC} may explain the discrepancy observed in $^{99}$Tc and $^{151}$Sm spectra but fail to reproduce the $^{63}$Ni and $^{241}$Pu spectra. 
Furthermore, their AQE correction overlooks the delicate interplay between atomic and nuclear matrix elements in forbidden transitions, which are encountered in all the aforementioned transitions except $^{3}$H and $^{63}$Ni.
Therefore, there are significant uncertainties in the theoretical predictions of AQE at very low energies.
A more accurate model, validated with high-precision measurements, is urgently needed.

\begin{figure*}[tb]
\centering
\includegraphics[width=.8\hsize]{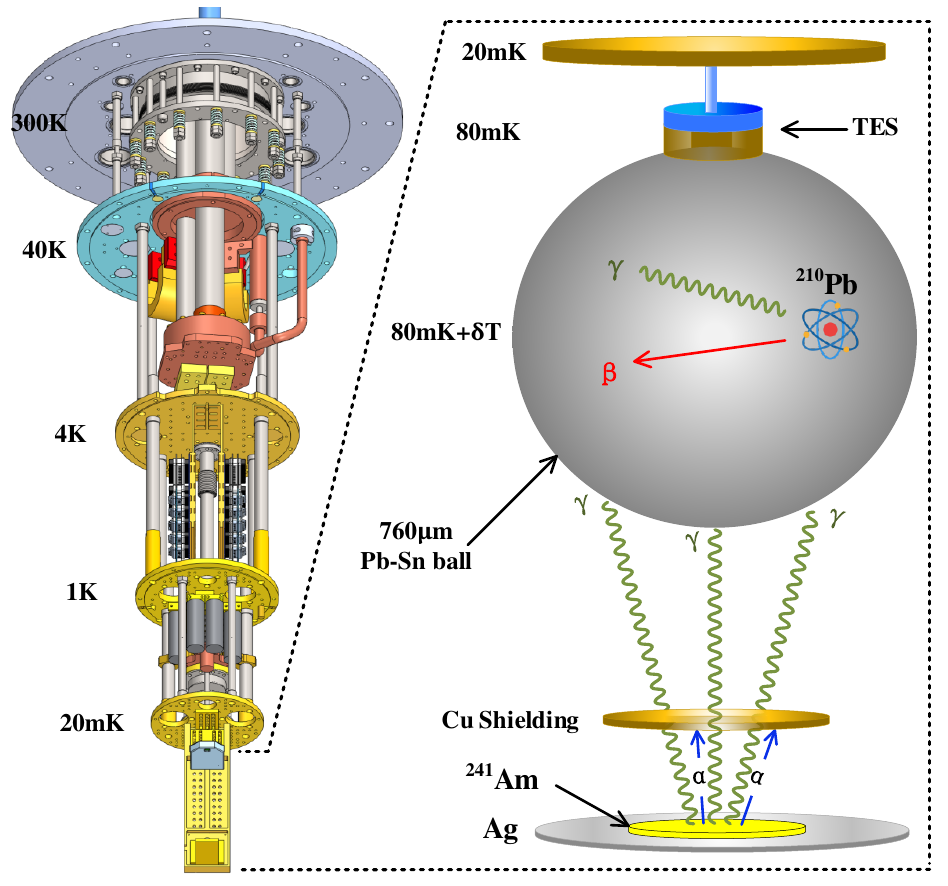}
\caption{Schematic drawing of the experimental setup.
The micro-calorimeter is operated in the dilution refrigerator at a base temperature of 20~mK. 
The Pb-Sn ball acts both as the $^{210}$Pb source and the energy absorber for the TES micro-calorimeter.
An $^{241}$Am $\gamma$ source is placed 22 cm from the absorber with a copper shielding in between to stop $\alpha$ particles from $^{241}$Am.}
\label{fig:setup}
\end{figure*}

\begin{figure*}[tb]
    \centering
\includegraphics[width=0.8\hsize]{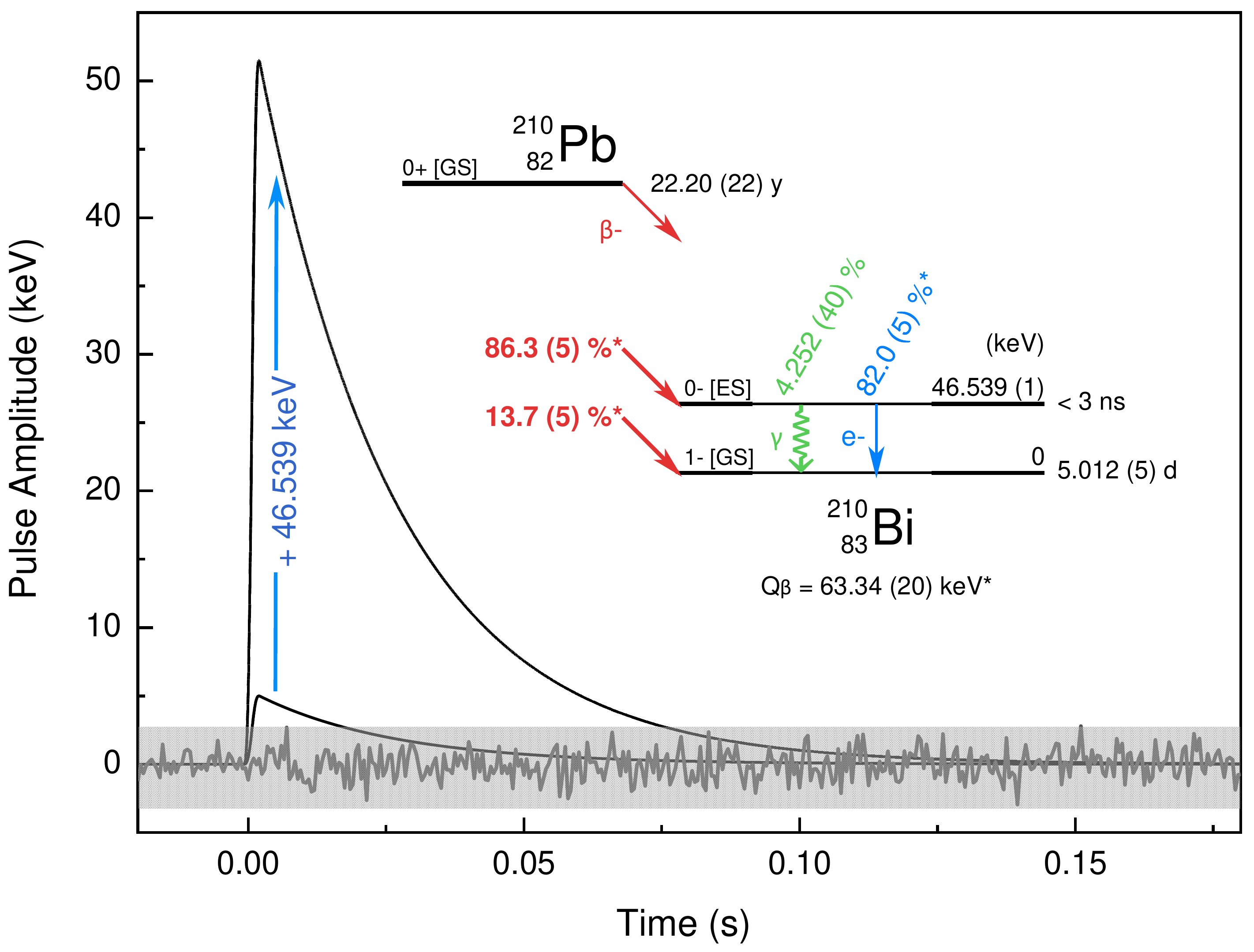}
\caption{Illustration of the threshold-free measurement. The inset presents the decay scheme of $^{210}$Pb disintegration based on the evaluated data from~\cite{ENSDF2014}, except for the quantities marked with an asterisk, which are from this study.
For most of the time, $^{210}$Pb decays to the excited state of $^{210}$Bi with an excitation energy $\Delta E = 46.539~(1)$~keV released in the form of internal conversion electrons or $\gamma$ rays within a time frame of nanoseconds.
In the micro-calorimeters, the $\beta$-ES pulse is shifted up because of $\Delta E$.
The shift would lift small $\beta$-ES signals away from the electronics and detector noise.
In the energy spectrum, the $\beta$-ES spectrum is shifted by $\Delta E$, from a range of [0, 17.0]~keV to [46.5, 63.5]~keV, enabling an entirely threshold-free measurement of the $\beta$-ES spectrum.}
\label{fig:coincidence}
\end{figure*}

%\section{Experimental setup}
\section*{The micro-calorimeter setup}

In this study, we present the first unambiguous precision measurement of a complete $\beta$ energy spectrum using a high-sensitivity, total-absorption Transition Edge Sensor (TES)-based micro-calorimeter.
The $^{210}$Pb decay scheme is dominated by the $\beta$ branch which goes to the first excited state (ES) of $^{210}$Bi with an excitation energy of $\Delta E =$ 46.539 (1)~keV\footnote{All the nuclear data used in the present study are from ENSDF evaluations~\cite{ENSDF2006,ENSDF2008,ENSDF2014}. All the Q-values are from AME2020 evaluations~\cite{Wang2021}.}.
%The $\beta$ transition energy is $17.0\pm 0.5$~keV, followed by a de-excitation from ES within 3~ns.
The micro-calorimeter measures the $\beta$ and subsequent de-excitation energy simultaneously, shifting the $\beta$-ES spectrum by $\Delta E$.
This shift enables direct measurement of AQE from 0 keV to the endpoint energy, circumventing energy detection thresholds entirely. 

The TES-based micro-calorimeter is thermally anchored on the mixing chamber plate of a dilution refrigerator, as shown in Fig.~\ref{fig:setup}. 
The system is cooled down to a base temperature of 20~mK, but the TES is heated to the superconducting transition edge ($\sim$80 mK), 
with stability maintained within 0.1\% by the bias current. 
A three-layer annealed $\mu$-metal shield reduces the ambient magnetic field to the 2~nT level and suppresses its temporal fluctuations by approximately three orders of magnitude, ensuring TES stability against spurious magnetic perturbations over an extended period.

The micro-calorimeter comprises a 760-$\mu$m-diameter lead-tin spherical absorber (63\% Sn, 37\% Pb by mass) coupled to the TES via epoxy. 
A picture of the micro-calorimeter can be seen in the Methods section. 
At 80~mK, energy deposition in the absorber induces a measurable temperature rise, which is detected by the TES and read out by a Superconducting Quantum Interference Device~(SQUID).
An $^{241}$Am source emitting 59.54092 (10) keV $\gamma$-rays is employed to illuminate the micro-calorimeter for energy calibration and stability correction. 
A 3.4~mm-thick copper shielding film is placed between the source and the spherical absorber to prevent $\alpha$ particles of $^{241}$Am from reaching the detector. 

Fig.~\ref{fig:coincidence} illustrates the threshold-free measurement enabled by the unique combination of $\beta$-ES decay of $^{210}$Pb and micro-calorimeter technique.
The Pb-Sn absorber contains a trace amount of $^{210}$Pb, which decays to the excited state of $^{210}$Bi with a reported branching ratio $B_{ES} =$ 86.3 (5)\%.
The excitation energy $\Delta E$ is released via internal conversion electrons or $\gamma$ rays within a time frame of nanoseconds, which is much shorter than the response time of a micro-calorimeter.
Therefore, the energies of the $\beta$ and de-excitation are measured simultaneously, with the $\beta$ pulse shifted by $\Delta E$, thereby lifting possibly small $\beta$ signals away from the electronics and detector noise.
In the energy spectrum, the $\beta$-ES spectrum is shifted by $\Delta E$, from [0, 17.0]~keV to [46.539, 63.5]~keV, enabling an entirely threshold-free measurement of the $\beta$-ES spectrum.

We conducted the measurement of $^{210}$Pb $\beta$-ES spectrum over 50 days. 
During the measurement, the system stability is actively monitored via the baseline fluctuations of the voltage signal from the SQUID.
The baseline drift typically remains below 10~mV, while the amplitude of the typical $\beta$-ES signal is on the order of 1~V.
The energy linearity calibration and temporal stability correction have been detailed in prior studies~\cite{2023-Shuo-NST}.

\section*{Theoretical $\beta$ spectra}

The analysis of our measurement requires an accurate description of the $\beta$ decay spectra of $^{210}$Bi and $^{210}$Pb. 
The former was obtained with the BetaShape code~\cite{2023-Mougeot-ARI}, which generates a spectrum based on the experimental shape factor~\cite{GRAUCARLES2005}. 
Our full \textit{ab initio} treatment based on the relativistic, low-energy effective theory~\cite{Behrens1971} accounts for lepton and nuclear currents, and atomic effects, which are required for modelling the two spectra in $^{210}$Pb first-forbidden, non-unique decays.
This effective theory, not being renormalizable, had radiative corrections from quantum electrodynamics added.

AQE is of utmost importance due to the very low energies of the $^{210}$Pb $\beta$ transitions. 
In particular, AQE encompasses the contribution of all atomic orbitals, from inner to outer shells (see Fig.~\ref{fig:AQE_correction} in Methods).
The required precise and consistent determination of the relativistic wave functions of the $\beta$ electron in the continuum and of the electrons of the $^{210}$Pb and $^{210}$Bi atoms was realized using a newly developed methodology designed for rigorous atomic modelling effects in electron capture. 
Atomic bound-wave functions were calculated using an optimized effective potential computed within the relativistic local density functional, and validated against state-of-the-art multi-configurational self-consistent field calculations over a wide range of atomic numbers~\cite{And24}. 
This approach was extended for the present study to include the calculation of the continuum $\beta$ electron wave functions within the same optimized effective potential, ensuring perfect consistency. 

The nuclear structure was determined from a realistic shell model~\cite{2014-Brown-nushellx} which provides one-body transition densities (OBTDs). 
These OBTDs enter as weights in a sum of nucleon-nucleon matrix elements that describes the $\beta$ transition. 
Atomic and nuclear wave functions were then accurately folded into the weak transition matrix elements. 
For comparison with our measurement, $\beta$ spectra with and without the AQE correction were produced. 
More details on our theoretical modelling are provided in the Methods section.

%\section{Results}
\section*{Combined analysis}
\label{sec:auto}
\begin{figure*}[tb]
\includegraphics[width=\hsize]{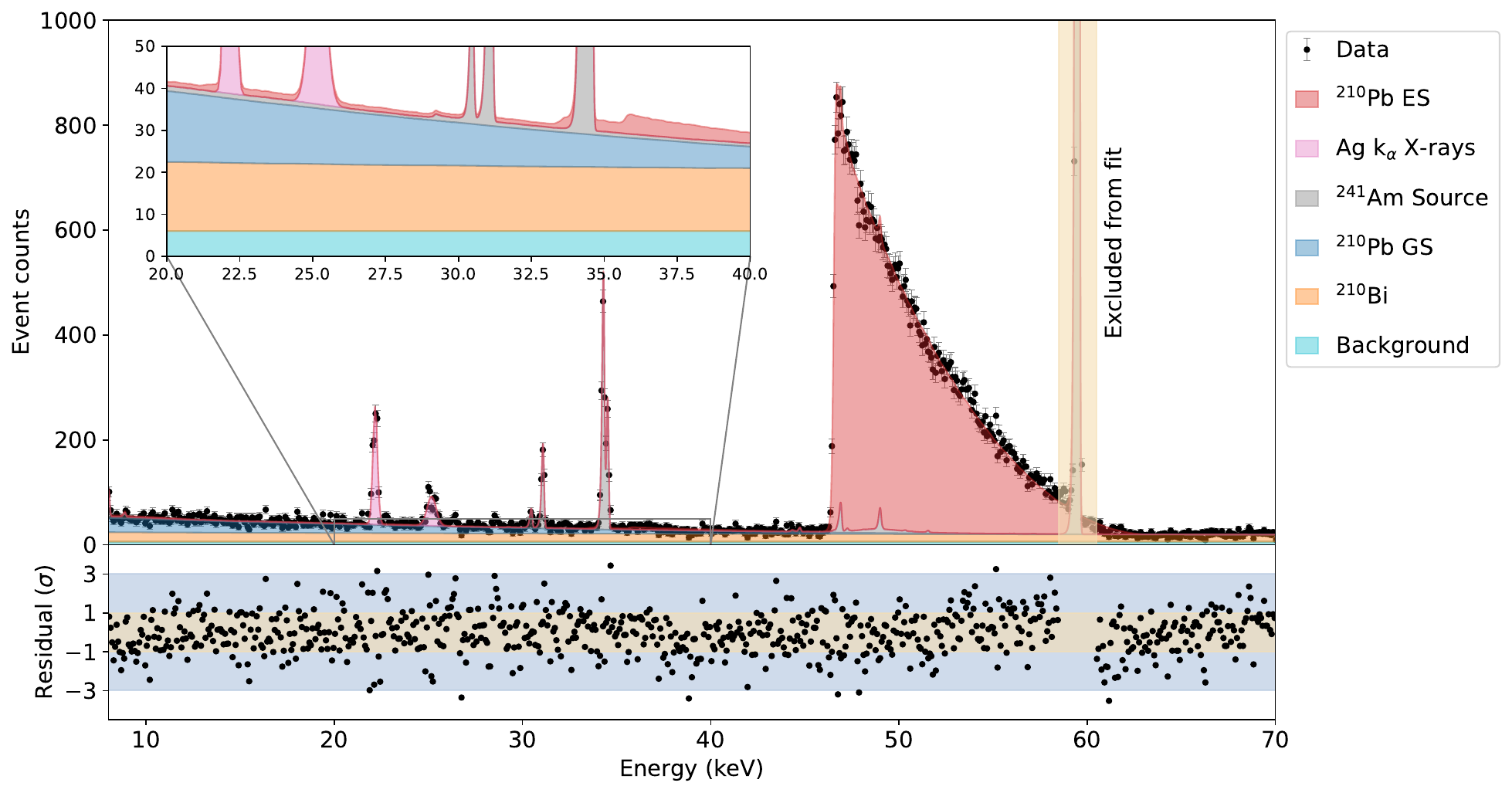}
\caption{The measured energy spectrum from 8 to 70~keV. 
The most prominent signal peak from 46.5 to 63.5~keV is the $\beta$-ES + $\Delta E$. 
The spectrum is fitted with the signal peak, Gaussian peaks of $K_\alpha$ of Ag, $^{241}$Am Compton continuum, and escape peaks in Sn, $^{210}$Pb $\beta$-GS, $^{210}$Bi $\beta$, and a flat background spectrum. 
The full absorption peak of $^{241}$Am is excluded from the fit.
The bottom panel shows the residual plot.
}
\label{fig:spectrumFull}
\end{figure*}

All the physical signals and backgrounds, including the theoretical $\beta$ spectra, were convoluted with the response function of the detection system from Geant4 simulations~\cite{Geant42003}.
The dominant effect came from the finite absorber size. 
A slight asymmetric line shape observed on the 59.5~keV calibration peak served as an additional detector response for convolution with the $\beta$-ES predicted spectrum.

The analysis and subsequent quantification of the AQE are sensitive to parameters of the $^{210}$Pb decay scheme. 
The $\beta$-GS spectrum acts as a background, thus rendering precise branching ratios particularly salient. 
The Q-value influences the spectrum shapes directly. 
In view of the inaccuracy of these quantities in the evaluations~\cite{ENSDF2014,Wang2021}, they were determined with improved uncertainties from the present measurement, and the decay scheme was re-evaluated, as detailed in Methods.
Parameters of the updated scheme were used for the current data analysis.

The contributions of the resulting spectra were fitted to the experimental data, along with background contributions from the Compton scattering of the 59.5~keV calibration peak, $^{210}$Pb $\beta$-GS spectrum, $^{210}$Bi $\beta$ spectrum, and a flat background. 
Decay chain backgrounds were constrained through temporal and spectral analysis. 
We correlated the event rate of $^{210}$Bi decay with $^{210}$Pb rate in the fit. 
Indeed, the $^{210}$Bi decay ($T_{1/2} =$ 5.012 (5) d, $Q_{\beta} =$ 1161.2 (8) keV) and the subsequent $^{210}$Po $\alpha$-decay ($T_{1/2} =$ 138.376 (2)~d, $Q_{\alpha} =$ 5407.53 (7) keV) have significantly shorter half-lives compared to $^{210}$Pb decay. 
Their decay rates are thus primarily governed by the latter. 
The Po-related $\alpha$ events caused detector saturation and were rejected via pulse shape discrimination.

The fit was performed from 8 to 70~keV within the measured energy range, as illustrated in Fig.~\ref{fig:spectrumFull}. 
The [58.5, 60.5]~keV range was excluded to prevent the calibration peak from dominating the fit statistic. 
The inset highlights the different contributions in addition to the $^{210}$Pb $\beta$-ES spectrum of interest.
Systematic uncertainties were evaluated through simulation, including the effects of absorber dimensions and decay data. 

\begin{figure*}[tb]
\centering
\includegraphics[width=0.8\hsize]{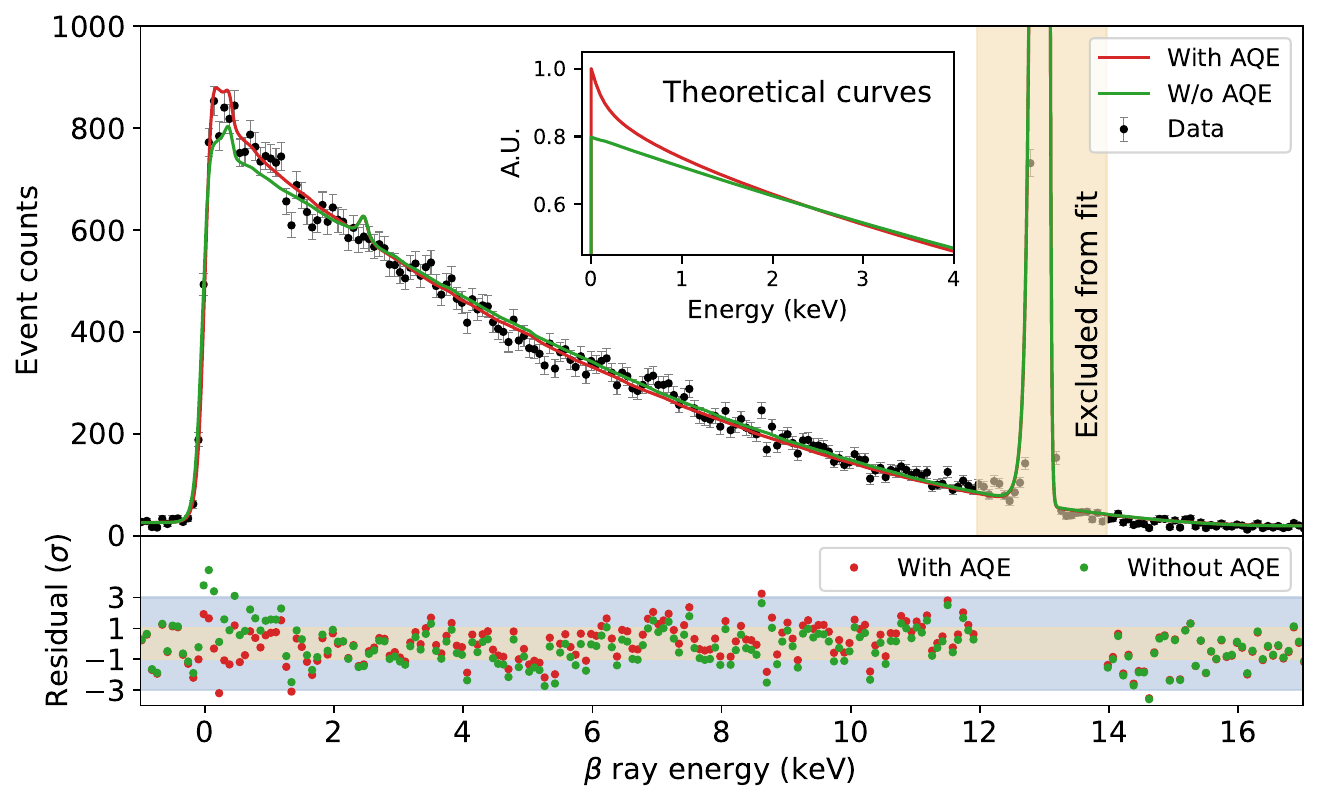}
\caption{The measured $\beta$-ES spectrum after shifting back to 0~keV starting point and theoretical $\beta$-ES calculations with (red curve) and without (green) AQE effect considered.
The spectrum bin width is 80~eV. 
The fit curves include the  $\beta$-ES signal and backgrounds, detailed in the appendix.
The bottom panel shows residuals between theories and data.
For each bin, the residual is calculated as (data-theory)/$\sqrt{\textrm{data}}$.
}
\label{fig:SpectrumES}
\end{figure*}

%\section{Summary and Outlook}
\section*{Quantification of AQE}

Fig.~\ref{fig:SpectrumES} compares the measured $\beta$-ES spectrum, subtracted by $\Delta E$, with the theoretical curves accounting for or not for AQE, whose amplitudes were left free in the fit.
The measured spectrum is unaffected by distortion of the spectral shape or threshold-related uncertainties in detection efficiency at null energy. 
The only factor affecting this region is smearing due to the energy resolution, which is 142~eV at the 59.54~keV peak of $^{241}$Am.
With the calibration peak excluded, the signal-to-background ratio of 10.2 shows that the systematics on the background have a minimal impact on the measurement.
The goodness-of-fit factor $\chi^2$ is 831 (883) for the fit with (without) AQE, with 748 degrees of freedom. 
The curve with AQE effect provides a better fit to the data with a statistical significance of 7.2 $\sigma$. 

To our knowledge, this study is the first high-precision measurement of the complete spectrum of a $\beta$ transition with no energy threshold, and the first full \textit{ab initio} calculation of its atomic and lepton components coupled with realistic nuclear structure.

All the atomic shells allowed by total angular momentum selection rules are demonstrated to participate in the increase of emission probability in the very-low-energy region, which can only be described with an accurate atomic model.
Since AQE is independent of the transition energy, our result strongly constrains the prediction of $^{212,214}$Pb $\beta$ spectra, which are key to background estimation and detection limits in dark matter experiments~\cite{LZ:2022lsv,PandaX:2024qfu,XENON:2023cxc}.
Furthermore, increased emission probability of low-energy electrons affects the spectral shape~\cite{Nitescu:2024tvj} of double beta decay~\cite{Dolinski:2019nrj} and limits the search for new physics signals~\cite{Deppisch:2020mxv}.
The entanglement of the electron and antineutrino emitted in $\beta$ decay implies that the increased emission probability at low energy $\beta$ occurs identically for antineutrinos at high energy, which may provide a possible explanation for recent reactor antineutrino anomalies~\cite{Boser:2019rta}. 
Our study demonstrates that increasingly precise experimental data can sharpen theoretical models within the SM, offering a pathway to explore physics beyond it.

\newpage
\section*{Methods}
\subsection*{Experiment}
\begin{figure*}[!htb]
\centering
\includegraphics[width=0.8\hsize]{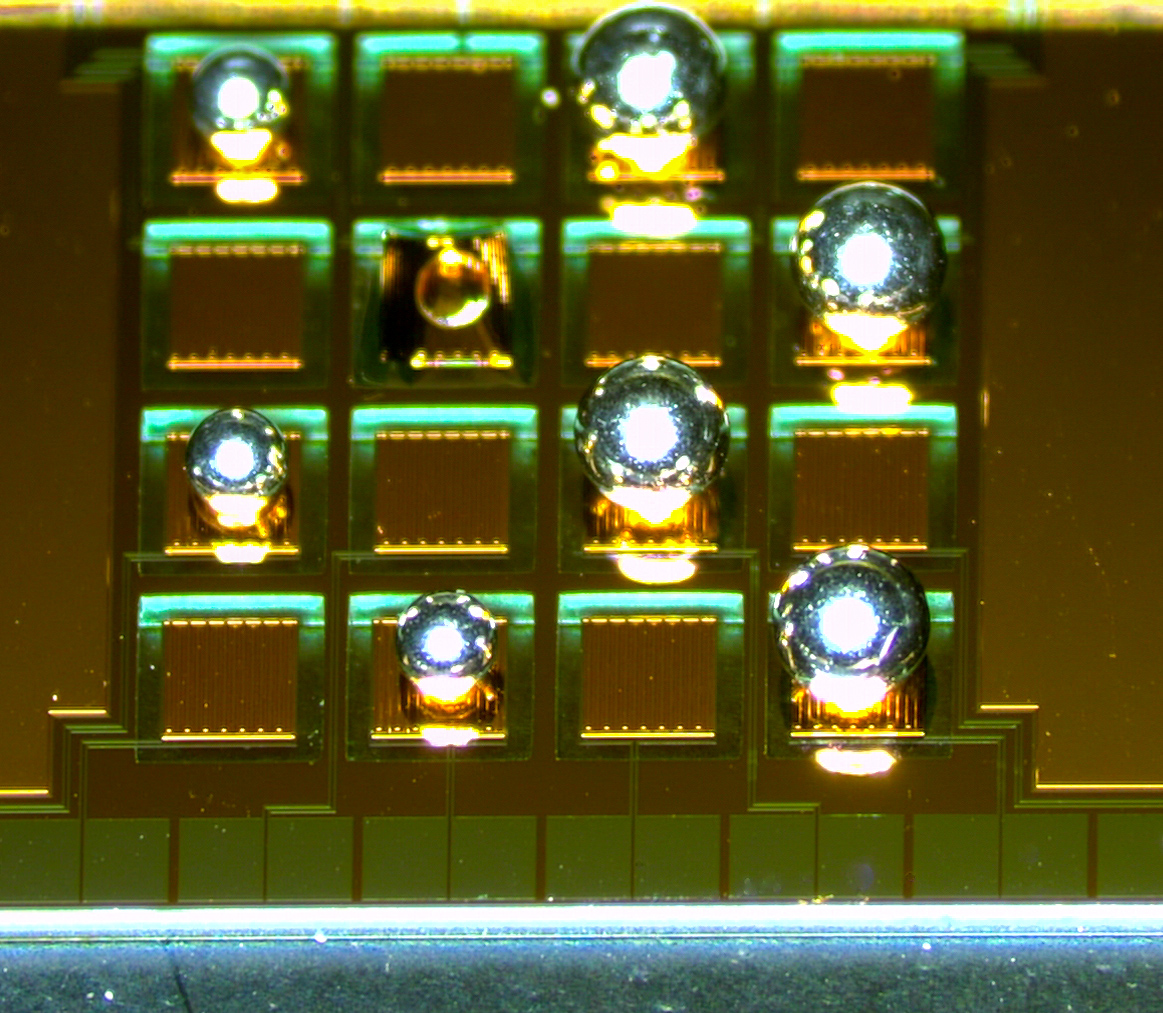}
\caption{The chip contains 16 TES pixels, each using a double-layer film of molybdenum and copper as a temperature sensor, and some of the pixels are bonded with lead-tin alloy balls. The outer ring of the chip is plated with a layer of gold to ensure optimal heat-conducting performance.}
\label{chipfig}
\end{figure*}

%\subsubsection*{Chips}\label{secA1}
%{\color{red}[XM: Explanations on chips must be rephrased properly. We never explain that several absorbers were measured and how the data acquired were summed up or selected.]}
Fig.~\ref{chipfig} shows a picture of the micro-calorimeter setup. 
A group of 16 TES chips is shown, while seven chips are instrumented with absorbers.
The absorbers of the micro-calorimeter setup were composed of 63\% tin and 37\% lead, with trace amounts of $^{210}$Pb present naturally. 
Three of the absorbers are 600~$\mu$m in diameter, and the other four are 760~$\mu$m in diameter. 
The absorber is thermally bonded to the TES chip via a 100-$\mu$m-diameter, 10-$\mu$m-thick Stycast 2850 epoxy layer, which simultaneously provides mechanical stability and optimized thermal conductance between the absorber and the TES.

With only one set of SQUID electronics, we tested all 7 setups in sequence.
The best-performing micro-calorimeter was selected, and we took data for 50 days to accumulate enough statistics. With the inclusion of the 59.54 keV gamma ray count for calibration, one pulse was recorded every 28 seconds; therefore, pile-up contribution in our measurement is totally negligible.

%\subsubsection*{Data process and spectrum calibration}\label{secA2}

The energy scale is calibrated using known peaks~\cite{Deslattes2003}: 21.99030 (10)~keV and 22.162917 (30)~keV (K$\alpha_2$ and K$\alpha_1$ Ag X-rays, respectively), 24.94242 (30)~keV (K$\beta_{1,3}$ Ag X-ray), 34.26956 (25)~keV and 34.49688 (25)~keV (K$\alpha_1$ and K$\alpha_2$ Sn X-ray escape peaks from $^{241}$Am $\gamma$-ray), and 59.54092 (10)~keV ($^{241}$Am $\gamma$-ray). 
The detector response is continuously monitored \emph{in situ} using the $^{241}$Am $\gamma$-ray. 
Throughout the data collection process, fluctuations in the temperature of the dilution refrigerator, TES, and absorbers may impact the detector response. 
The time series of the pulse amplitude $V_n(t_n)$ of the 59.54 keV $\gamma$ events is smoothed using a LOWESS (locally weighted scatterplot smoothing) technique to obtain the empirical correlation function ${S}_n(t_n)$, which is then used to correct the detector response. 
This correction improves the energy resolution from 181~eV to 142~eV at 59.54 keV. 
A polynomial function is used to fit the pulse amplitude with the afore-mentioned peaks for the energy calibration. 

\subsection*{Uniqueness of $^{210}$Pb $\beta$ decay}

$^{210}$Pb is an ideal choice for a threshold-free measurement of $\beta$ spectrum.
The $\beta$ energies and de-excitation energy are both $\mathcal{O}$(10~keV) and in the optimal response range of our micro-calorimeter. 
Most of the time, the energies are fully absorbed in our absorber.
The probability of $\beta$-ES particles and internal conversion electrons escaping the absorber is less than $10^{-4}$, and thus negligible in our measurement.
Monte Carlo simulations indicate approximately 20\% probability in the 46.539 keV $\gamma$-rays escape from the 760-$\mu$m absorber.
However, given the 4.252 (40)\% $\gamma$-ray emission probability for the de-excitation, fewer than 1\% of decay events exhibit undetected $\Delta E$. 
These $\gamma$-escape events produce reconstructed energies below the ROI and do not contribute to the measured $\beta$-ES spectrum.
The branching ratio to the ES is more than five times the ratio to the ground state (GS), indicating that the $\beta$-ES spectrum is the dominating feature in our measurement. 
The subsequent $^{210}$Bi $\beta$ spectrum spreads from null energy to 1161.2 (8) keV with a single $\beta$-GS branch, and its impact on the ROI is minimal, as well.
We have simulated the detector response to both the ES and GS $^{210}$Pb decays, as well as to the $^{210}$Bi decay, using the Geant4 framework~\cite{Geant42003}. More details can be found in the Methods/Experiment.
It is also worth noting that the half-life of $^{210}$Pb ($T_{1/2} =$ 22.20 (22)~y) also favours our setup. 
A trace amount of $^{210}$Pb occurs naturally, so that the sample preparation is straightforward, and also provides a sufficient number of events within a relatively short data-taking time. 

\subsection*{Theory}
The $^{210}$Pb $\beta$ decays are of first forbidden non-unique nature. 
The nuclear structure used in the spectrum calculation and the methodology are detailed in~\cite{2022-Karsten-ARI,Pau24}. 
The NushellX@MSU shell model code~\cite{2014-Brown-nushellx} was used to determine the list of single-particle transitions, weighted by their one-body transition densities, describing each of the two $^{210}$Pb transitions between the involved nuclear states. 
Above a doubly magic $^{208}$Pb core, we selected the \textit{jj67pn} valence space without any restriction and the recommended \textit{khpe} effective interaction~\cite{1991-Warburton-khpe}. The $\beta$-decay modelling requires relativistic nucleon wave functions: the large component is identified with the non-relativistic harmonic oscillator, and the small component is deduced in the non-relativistic limit; the total wave function is then renormalized. The ES transition of interest is dominated at $\gtrsim$ 95\% by the $n(2g_{9/2}) \longrightarrow p(1h_{9/2})$ nucleon-nucleon transition and involves only axial matrix elements of rank 0.
The small number of matrix elements involved, due to minimal total angular momentum change, and the very low transition energy significantly hinder the influence of nuclear structure on the $\beta$ spectrum shape, which influence is $\lesssim$ 0.5\% above 250 eV, increasing up to 1.7\% down to 100 eV, and to 3\% below.
%Such a small influence is expected because of the very low transition energies and the number of matrix elements limited by the total angular momentum change.
The GS transition is slightly more complicated but is still dominated at $\sim$ 82\% by the same nucleon-nucleon transition, and involves both vector and axial matrix elements of rank 1. Relativistic vector matrix elements are determined assuming the Conserved Vector Current (CVC) hypothesis. Influence of an effective $g_A$ value is inexistent on the ES spectrum shape and was found $\lesssim$ 0.05\% on the GS spectrum shape.

\begin{table}[!htb]
%\caption{Comparison of theoretical and experimental atomic orbital energies. Measurements are from Bearden and Burr~\cite{Bearden67} except otherwise stated.}\label{tab1}
\caption{Comparison of theoretical and experimental atomic orbital energies. Measured values are from the compilation of Sevier~\cite{Sevier79}, except the ionization energies $(^{\dag})$ which are from the NIST database~\cite{NIST24}.}\label{tab1}
\begin{tabular*}{\textwidth}{@{\extracolsep\fill}ccccccc}
\toprule%
& \multicolumn{3}{@{}c@{}}{$^{210}$Pb} & \multicolumn{3}{@{}c@{}}{$^{210}$Bi} \\\cmidrule{2-4}\cmidrule{5-7}%
Orbital & E$_{\text{theo}}$ (eV) & E$_{\text{exp}}$ (eV) & $|\Delta|$ (\%) & E$_{\text{theo}}$ (eV) & E$_{\text{exp}}$ (eV) & $|\Delta|$ (\%) \\
\midrule
$1s_{1/2}$&	-88071.47	&-88004.5 $\pm$ 0.7	& 0.08   &-90600.81	&-90525.9 $\pm$	0.7	&0.08 \\
$2s_{1/2}$&	-15786.23	&-15860.8 $\pm$	0.5	&0.47   &-16312.07	&-16387.5 $\pm$	0.5	&0.46 \\
$2p_{1/2}$&	-15197.81	&-15200.0	$\pm$ 0.4	&0.01   &-15711.36	&-15711.1 $\pm$	0.4	&0.00 \\
$2p_{3/2}$&	-12994.86	&-13035.2 $\pm$	0.3	&0.31   &-13378.13	&-13418.6 $\pm$	0.3	&0.30 \\
$3s_{1/2}$&	-3803.99	&-3850.7 $\pm$	0.5	&1.21   &-3952.39	&-3999.1 $\pm$	0.5	&1.17 \\
$3p_{1/2}$&	-3531.51	&-3554.2 $\pm$	0.3	&0.64   &-3673.81	&-3696.3 $\pm$	0.3	&0.61 \\
$3p_{3/2}$&	-3038.85	&-3066.4 $\pm$	0.4	&0.90   &-3150.20	&-3176.9 $\pm$	0.4	&0.84 \\
$3d_{3/2}$&	-2579.24	&-2585.6 $\pm$	0.3	&0.25   &-2681.64	&-2687.6 $\pm$	0.3	&0.22 \\
$3d_{5/2}$&	-2473.88	&-2484.0 $\pm$	0.3	&0.41   &-2570.08	&-2579.6 $\pm$	0.3	&0.37 \\
$4s_{1/2}$&	-867.04	&-893.6 $\pm$	0.7	&2.97   &-913.23	&-938.2 $\pm$	0.7	&2.66 \\
$4p_{1/2}$&	-747.75	&-763.9 $\pm$	0.8	&2.11   &-790.67	&-805.3 $\pm$	0.8	&1.82 \\
$4p_{3/2}$&	-626.71	&-644.5 $\pm$	0.6	&2.76   &-661.17	&-678.9 $\pm$	0.6	&2.61 \\
$4d_{3/2}$&	-428.95	&-435.2 $\pm$	0.5	&1.44   &-458.27	&-463.6 $\pm$	0.5	&1.15 \\
$4d_{5/2}$&	-406.22	&-412.9 $\pm$	0.6	&1.62   &-433.96	&-440.0 $\pm$	0.6	&1.37 \\
$4f_{5/2}$&	-148.58	&-148.5 $\pm$	0.1	&0.05    &-169.02	&-162.4 $\pm$	0.1	&4.08 \\
$4f_{7/2}$&	-143.38	&-143.6 $\pm$	0.1	&0.16   &-163.35	&-164.9 $\pm$	0.1	&0.94 \\
$5s_{1/2}$&	-151.81	&-147.3 $\pm$	0.8	&3.06    &-165.65	&-159.3 $\pm$	0.8	&3.98 \\
$5p_{1/2}$&	-110.22	&-104.8 $\pm$	1.0	&5.17    &-122.26	&-116.8 $\pm$	1.0	&4.68 \\
$5p_{3/2}$&	-86.715	&-86.0 $\pm$	1.0 &0.83    &-96.395	&-92.8 $\pm$	1.0	&3.87 \\
$5d_{3/2}$&	-28.724	&-28.25 $\pm$ 0.05	&1.68    &-35.155	&-26.5 $\pm$	0.05 $^*$&32.7 \\
$5d_{5/2}$&	-25.958	&-25.28 $\pm$ 0.05	&2.68    &-31.959	&-24.4 $\pm$	0.06 $^*$&31.0 \\
$6s_{1/2}$&	-16.652	& -- & -- &-19.195	& -- & -- \\
$6p_{1/2}$&	-8.6488	& -- & -- &-10.221	& -- & -- \\
$6p_{3/2}$&	-6.8128		&-7.4166799 $\pm$ 0.0000006 $^{\dag}$ &8.14 &  -7.9143	&-7.285516 $\pm$ 0.000006 $^{\dag}$& 8.63 \\
\botrule
\end{tabular*}
\footnotetext{$^*$ These Bi orbitals should be more bound than those of Pb. The difference is due to different chemical states in the measurements.}
%\footnotetext{$^{\dag}$ Ionization energy from NIST database~\cite{NIST24}.}
\end{table}

The $\beta$ electron emitted during the $^{210}$Pb decay then passes through the electron cloud of the $^{210}$Bi daughter atom and can be exchanged with one of the bound electrons of the various shells. Due to total angular momentum selection rules, only the $s_{1/2}$ and $p_{1/2}$ shells can be involved in this process. In this view, we use the sudden approximation, which consists of assuming that the time for the transmutation of the nucleus is instantaneous compared with the time for the rearrangement of the electrons in the daughter atom. Therefore, as the $\beta$ electron passes through the electron cloud, it feels the potential of the daughter atom, a nucleus with charge $Z=83$ but with $82$ bound electrons. Another constraint imposed by the physics of the process is that asymptotically, i.e., far from the nucleus, the $\beta$ electron must experience a Coulomb potential of the form $(+1/r)$. Furthermore, a fully consistent atomic model requires the wave functions of the bound and continuum states to be calculated using the same Hamiltonian, all of them being orthogonal to each other by construction. Finally, especially in determining the bound-state energy levels of the parent and daughter atoms, the atomic model must be as accurate as possible. 

In a recent study~\cite{And24}, we reported on a newly developed atomic model based on the relativistic local density approximation (RLDA) that perfectly satisfies these requirements. It includes correlation effects and a precise description of the nucleus, which is required for heavy atoms where the electron wave functions of the inner shells penetrate inside the nucleus. In addition, this model employs a method (called KLI) based on an optimized effective potential that was developed in a non-relativistic framework by Krieger et al.~\cite{Krieger92,Li93} and extended to the relativistic case by Tong and Chu~\cite{Tong98}. Table~\ref{tab1} illustrates the high accuracy of our predictions by comparing the theoretical orbital energies to experimental values. This approach has been extended for this work to include the calculation of the continuum $\beta$ electron wave functions within the same optimized effective potential, ensuring perfect consistency. 

\begin{figure*}[!htb]
\centering
\includegraphics[width=0.8\hsize]{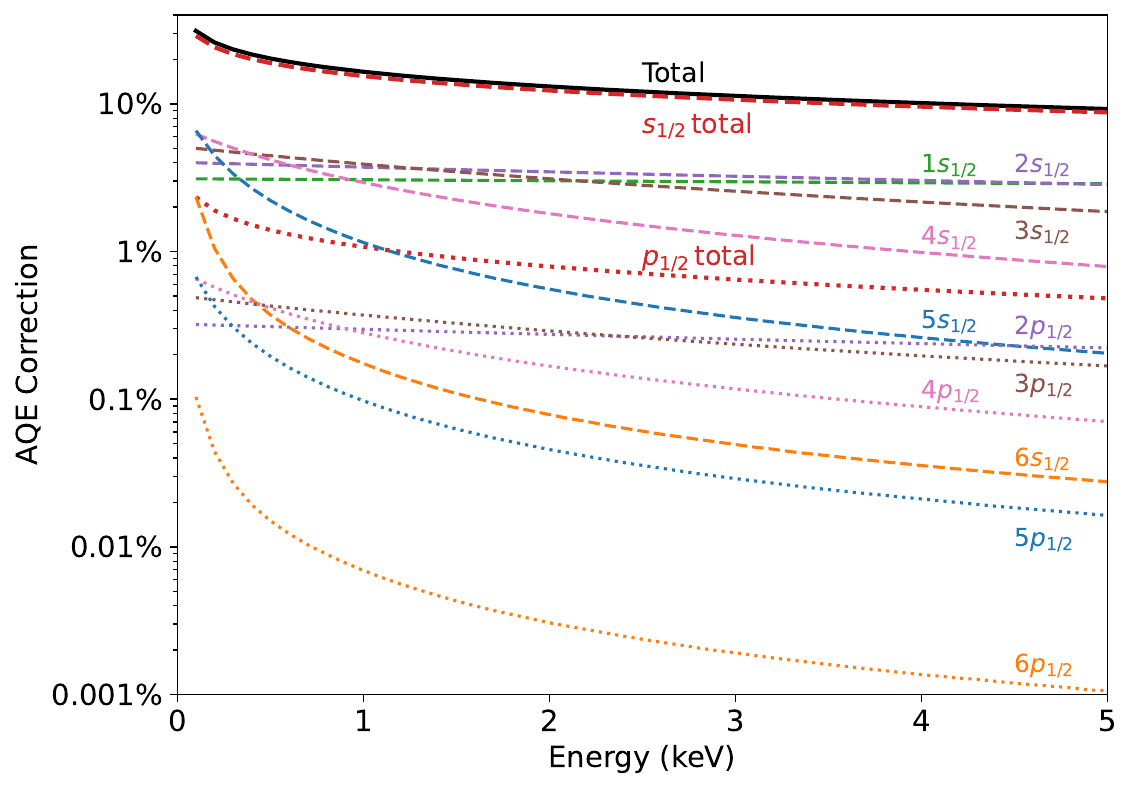}
\caption{Total atomic quantum exchange (AQE) correction (in percent), the total contributions of the $s_{1/2}$ and $p_{1/2}$ orbitals, and the contributions of each subshell. The $s_{1/2}$ orbitals predominantly contribute the total AQE correction.}
\label{fig:AQE_correction}
\end{figure*}

In~\cite{2023-Mougeot-ARI}, we presented an extension of the AQE formalism to forbidden transitions, focusing on the unique ones. The first steps of the derivation introduce the atomic wave functions into the lepton current, providing the basis for calculating forbidden non-unique transitions. In the present study, we introduced the AQE following the same method, which included proper convolution with the nuclear current. This allowed us to calculate the $\beta$ spectrum comprehensively. Fig.~\ref{fig:AQE_correction} presents the total AQE correction and its breakdown by atomic orbitals in the low-energy region of the $\beta$ spectrum. If the global amplitude of the correction is due to the innermost shells, it is clear that all the orbitals up to the valence shell significantly contribute to the shape of the AQE, and thus of the $\beta$ spectrum, below 1~keV.

%{\color{red}\subsection*{Determination of decay scheme parameters}
\subsection*{Determination of decay scheme parameters}
The most recent evaluation has yielded branching ratios for the $^{210}$Pb decay of $B_{GS} =$ 16 (3)\% and $B_{ES} =$ 84 (3)\%~\cite{ENSDF2014}. It was observed that they were not sufficiently accurate to constrain the global fit of our analysis. Consequently, they were permitted to vary during the fitting process. Systematic uncertainties were evaluated through simulations, including the effects of absorber dimensions. The resulting values were determined to be $B_{GS} =$ 13.7 (5)\% and $B_{ES} =$ 86.3 (5)\%, representing the most precise determination of the branching ratios to date, with an uncertainty improved by a factor of six.

The Q-value of the $^{210}$Pb decay was determined by following the method detailed in~\cite{Pau24}. The Kurie plot of the $\beta$-ES spectrum was fitted according to various assumptions regarding the calculation. The reference value was determined by assuming a numerical Fermi function, incorporating atomic screening and spatial extension of the nucleus, as well as atomic exchange and overlap corrections. The derivation of the radiative correction assumes that the emitted soft photons are not detected, while being fully reabsorbed in our detection system. It was thus not included, but considering this correction provided a conservative uncertainty component. The estimation of other components was achieved by varying the assumptions of the calculation, the energy range of the fit and the background contribution. The uncertainty budget is dominated by the statistics, with 0.16 from the fit method, 0.12 from the background and 0.011 from the theoretical modelling.  The maximum energy of the $\beta$-ES spectrum was determined to be E$_{\mathrm{max}} = $ 16.80 (20) keV. Combining with the level energy of 46.539 (1)~keV from~\cite{ENSDF2014}, the Q-value established from our measurement is $Q_{\beta} = $ 63.34 (20) keV. This outcome is consistent with the evaluated AME2020 value of 63.5 (5) keV~\cite{Wang2021}.

The decay scheme was then revisited based on these findings. In the most recent evaluation~\cite{ENSDF2014}, the probability of the $\gamma$ emission has been evaluated from measurements. The $\beta$ branching ratios have then been deduced with the assistance of the total internal conversion coefficient $\alpha_T$, calculated with the BrIcc code~\cite{Kibedi08} assuming a pure M1 transition. 
Maintaining the same $\gamma$ emission probability and utilizing the branching ratios determined in the present study, the conversion electron emission probability was calculated to be P$_{ce} =$ 82.0 (5)\%, a figure independent of $\alpha_T$ and with an uncertainty 2.6 times smaller. It was only possible to achieve a fully consistent decay scheme by adjusting the mixing ratio $\delta$ of the $\gamma$ transition. The adjustment was made using a series of BrIcc calculations. This result demonstrates the existence of a slight 0.636 (16)\% E2 admixture in the M1 $\gamma$ transition, with $|\delta| =$ 0.080 (1) and $\alpha_T =$ 19.30 (22).

\providecommand{\noopsort}[1]{}\providecommand{\singleletter}[1]{#1}%

\section*{Competing interests}
The authors declare no competing interests.

\section*{Acknowledgements}

We thank Dan McCammon, He Xiaotao, He Qinghua, Yao Jiangming, Dai Xiongxin, Zhang Guoqiang, Ma Long, Yang Xiaofei, and Jorn Beyer for discussions. This work is supported by the National Natural Science Foundation of China under the National Major Scientific Research Instrument Development Project (Grant No. 11927805). The theoretical work is conducted as part of the EMPIR Project 20FUN04 PrimA-LTD, which has received funding from the EMPIR program co-financed by the participating states and the European Union's Horizon 2020 research and innovation program.

\section*{Author contributions}

S.Z., K.H., and X.M. conceived the project.
S.Z. conducted detector setup, instrumentation, and data collection with assistance from T.S., W.-T.W., R.C., X.-P.Z., F.-Y.F., J.-C.L., J.-K.X., and Z.L..
K.H. and H.-R.L. analysed data with assistance from M.-Y.G..
X.M., P.-A.H., and A.A. performed theoretical calculations.  
S.Z., H.-R.L., K.H., X.M., and P.-A.H. drafted the manuscript with assistance from L.Z..
All authors discussed the results and contributed to the paper.

\section*{Materials \& Correspondence}
Correspondence and requests for materials should be addressed to ke.han@sjtu.edu.cn (K.H.), xavier.mougeot@cea.fr (X.M.), and paul-antoine.hervieux@ipcms.unistra.fr (P.-A.H.)

% \section*{Reprints}
% Reprints and permissions information is available at www.nature.com/reprints.
%\section*{Declarations}
% XXX':

% \begin{itemize}
% \item Conflict of interest/Competing interests (check journal-specific guidelines for which heading to use)
% \item Ethics approval and consent to participate
% \item Consent for publication
% \item Materials availability
% \item 
% \end{itemize}

\end{document}